\documentclass[twoside,final]{pro12}

\usepackage[ansinew]{inputenc}
\usepackage{amsmath,url}
\usepackage{tabularx,multirow,hhline}
\usepackage{graphicx} 

\head{B. Cecconi et al.} {Refurbishing Voyager/PRA data}	

\begin{document}

\title{REFURBISHING VOYAGER 1 \& 2 PLANETARY RADIO ASTRONOMY (PRA) DATA}	
\author{B. Cecconi\adress{\textsl LESIA, Observatoire de Paris, PSL Research University, CNRS, Sorbonne Universit\'es, UPMC Univ. Paris 06, Univ. Paris Diderot, Sorbonne Paris Cit\'e, Meudon, France}$\,$, A. Pruvot$^*$, L. Lamy$^*$, P. Zarka$^*$, C. Louis$^*$, \\
S. L. G. Hess\adress{\textsl ONERA, Toulouse, France}$\,$, D. R. Evans\adress{\textsl Radiophysics Inc.\ (retired), Boulder, CO, USA}$\,$, D. Boucon\adress{\textsl CST, CNES, Toulouse, France}}

\maketitle

\begin{abstract}
Voyager/PRA (Planetary Radio Astronomy) data from digitized tapes archived at CNES have been reprocessed and recalibrated. The data cover the Jupiter and Saturn flybys of both Voyager probes. We have also reconstructed goniopolarimetric datasets (flux and polarization) at full resolution. These datasets are currently not available to the scientific community, but they are of primary interest for the analysis of the Cassini data at Saturn, and the Juno data at Jupiter, as well as for the preparation of the JUICE mission. We present the first results derived from the re-analysis of this dataset.\end{abstract}

\section{Introduction}
The Planetary Radio Astronomy (PRA) experiments [Warwick et al., 1977] onboard the Voyager 1 and 2 spacecraft were the first radio instruments to explore the low frequency radio emissions of the giant planets. The PRA experiment observes radio waves up to 40.5~MHz, and is capable of measuring the sense of circular polarization of the observed radio waves. 

The Voyager 1 spacecraft approached Jupiter in March 1979 and Saturn in November 1980. The Voyager 2 spacecraft approached Jupiter in July 1979, Saturn in August 1981, Uranus in January 1986 and Neptune in August 1989. During each planetary flyby intense radio emissions of auroral origin were observed [Warwick et al, 1979a, 1979b, 1981, 1982, 1986, 1989]. The Jovian radio emissions reach 40 MHz, as discovered from the ground [Burke and Franklin, 1955], whereas all other giant planets show radio emission up to $\sim$1~MHz, as in the case of terrestrial auroral radio emissions [Zarka, 1998]. Atmospheric radio flashes (lightning electrostatic discharges) were also observed at Saturn, Uranus and Neptune [Zarka and Pedersen, 1983, 1986].

The PRA data were analyzed by the Voyager team and were partly archived at the NASA/PDS (Planetary Data System) PPI node (Planetary Plasma Interactions). Two PRA datasets are available in the archive: the full resolution Low Frequency band (LF) data\footnote{\texttt{VG1-J-PRA-3-RDR-LOWBAND-6SEC-V1.0} and \texttt{VG2-J-PRA-3-RDR-LOWBAND-6SEC-V1.0}: Voyager 1 and 2 PRA data during the Jupiter flyby dataset, available at NASA/PDS-PPI.} covering 1.2~kHz to 1.3~MHz, and 48~s average spectra High Frequency band (HF) data\footnote{\texttt{VG1-J-PRA-4-SUMM-BROWSE-48SEC-V1.0} and \texttt{VG2-J-PRA-4-SUMM-BROWSE-48SEC-V1.0}: Voyager 1 and 2 PRA data during the Jupiter flyby dataset, available at NASA/PDS-PPI.} covering 1.2~MHz to 40.5~MHz.

While managing ``orphaned'' datasets in its repository, the CNES long term archive service (SERAD, \textsl{Service de R\'ef\'erencement et d'Archivage des Donn\'ees}, Data Referencing and Archiving Service) identified a series of digitized Voyager PRA magnetic tape dumps. This paper describes the refurbishment of this dataset.

\section{PRA Instrument}
\begin{figure}
\centering\includegraphics[width=0.7\linewidth]{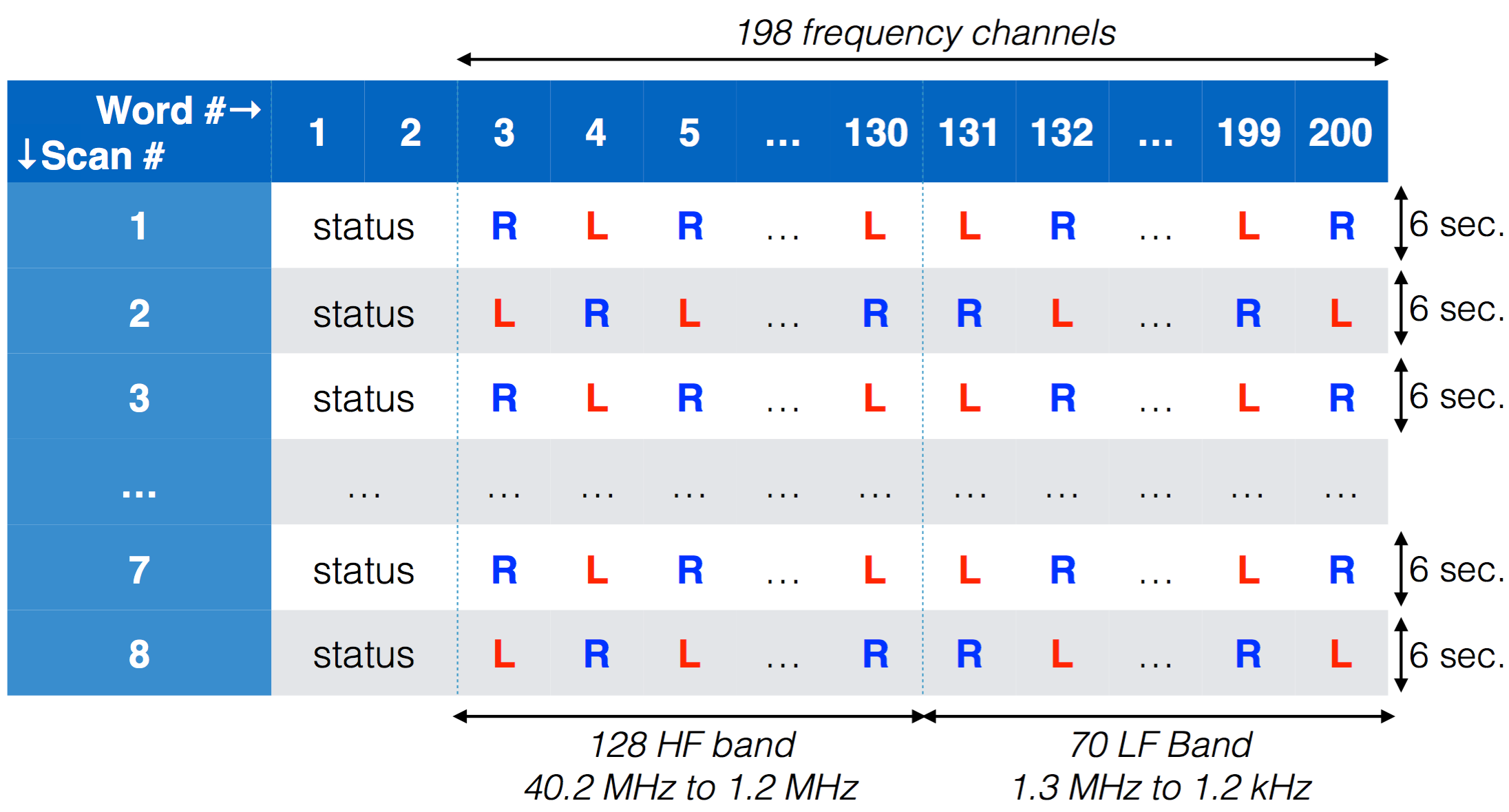}
\caption{Voyager PRA sampling scheme in its ``POLLO'' mode. Each DEDR ``record'' is composed of 8 ``scans'' containing 200 data ``words'' (16 bits values). The two first words of each scan contain a 32 bits status, followed by 198 measurement words, 128 for High Band (HB) and 70 for Low Band (LB), with alternate antenna combination ($R=V_+$ and $L=V_-$)}
\label{fig1}
\end{figure}

The PRA experiment is composed of a radio receiver sampling radio electric signals from 1.2~kHz to 40.5~MHz, and connected to a pair of orthogonal 10~m electric monopoles. The receiver is divided into two bands: the low band (LB) from 1.2~kHz to 1326~kHz composed of 70 spectral bins of 1~kHz every 19.2~kHz; the high band (HB) from 1.2~MHz to 40.2~MHz composed of 128 spectral bins of 200~kHz every 307.2~kHz. The radio signals sensed by the pair of monopoles are combined with a positive or negative phase quadrature: $V_\pm = V_1 \pm iV_2$, where $V_1$ and $V_2$ are the voltage sensed at antenna 1 and 2 respectively.
In its ``POLLO'' mode (the main radio astronomy mode) the receiver sweeps the HB and LB bands from high to low frequencies. The receiver measures the autocorrelation ($A_\pm = \left<V_\pm.V^*_\pm\right>)$ switching the antenna voltage combination every other measurement (see Fig.\ \ref{fig1}), so that polarization measurements can be derived. The autocorrelations can be combined to retrieve the spectral flux density and the circular polarization of the observed radio waves as $(A_+ + A_-)/2 \propto S$ and $(A_+ - A_-)/(A_+ + A_-) \propto V$, where S and V are the flux density and the circular polarization degree (details are given in section \ref{sec:polar}). Since this derivation requires an assumption on the radio wave's direction of arrival, it is a genuine goniopolarimetric inversion as defined in [Cecconi, 2004; 2007]. If the radio wave is circularly polarized, the proportionality factors only depend on the geometry of observation (orientation of the antenna system with respect to the wave direction of propagation). 

\section{Data format}
The magnetic tape dumps available at CNES/SERAD are composed of two high resolution datasets: EDR (Experiment Data Records) and DEDR (Decalibrated\footnote{In this context, ``decalibrated'' means that a calibration has been applied to the data. This misleading naming convention is not used in current data processing pipelines.} Experiment Data Records). The DEDR dataset contains calibrated data and includes data from both Voyager spacecraft for both Jupiter and Saturn flybys. The EDR dataset contains uncalibrated data and covers all planetary flybys. The SERAD archive contains 110 DEDR files and 1255 EDR files. The present paper deals with the DEDR dataset. Further work is planned to process the EDR dataset.

The documentation attached to the SERAD datasets were limited to a technical description of the data stream in the tapes: the DEDR science record data format and EDR record data format specification [pages extracted from Street, 1977] and the archive description and inventory file [Huc, 1996]. The organization of the EDR header, used for EDR and DEDR tapes, is described, as well as that of the DEDR format. Information on the operating mode was coded in the mode configuration words. Other documents are available online and described the overall spacecraft and instrument configurations, modes and characteristics. Limited documentation on the calibration (e.g., EDR to DEDR data level) was also available in the laboratory since a few researchers of the team worked on Voyager-PRA data when the instrument was active. Finally, thanks to D.R.\ Evans (retired from Radiophysics Inc.) personal cheat-sheet, we could decode the observation mode configuration and have access to the complete configuration of PRA for each record.

The DEDR record data section is described in [Street, 1977] and is recalled in Fig.\ \ref{fig1}. Each spectral sweep (i.e., ``scan'') lasts for 6~s. Individual measurements are made every 30~ms. The DEDR data files contain calibrated data. The calibrated ``POLLO'' (Polarization Low rate) data have been extracted and reordered into daily files. This dataset is named L2. 

\section{Polarization derivation}
\label{sec:polar}
The full resolution PRA data can be processed to extract flux density and polarization measurements, combining $A_+$ and $A_-$ data samples. Due to the sampling scheme, this  combination can be done two ways: inversion L3f uses $A_\pm$ samples at the same frequency on two successive scans; inversion L3t is using successive $A_\pm$ samples, from adjacent frequency bins. Inversion L3f has better spectral resolution, but requires consistent signal for 6~s. Inversion L3t has better temporal resolution (signal must be stable over 30~ms), but mixes signals from different spectral bins. Both inversions requires an assumption on the direction of arrival of the radio wave. Similarly to the ``two-antenna polarimeter'' inversion of Cecconi and Zarka [2005] (see section 2.1.3.1 in that paper), we assume here that the radio wave is propagating on the direction define by the spacecraft to observed planet line. 

Both L3f and L3t inversions require an accurate antenna calibrations. Several antenna calibration for the PRA experiment have been published [Warwick et al, 1977; Ortega-Molina and Lecacheux, 1990; Wang and Carr, 1994] but are not fully consistent one to the other, neither with notations nor results. The analysis and selection of the antenna calibrations will be conducted in the next months. 

Following the same formalism as [Cecconi, 2005], in the so-called ``short antenna range'' (wavelength is large compared to the antenna length), the voltage at the antenna feed is $V_i = \vec{E}.\vec{h}_i$. This scalar product is valid in any frame, but the wave frame is more suitable. The wave frame is defined such that $\vec{z}$ is the wave vector and the $\vec{x}$ and $\vec{y}$ axes complete the orthogonal triad. In this frame, the electric field $\vec{E}$ has only two components and the scalar product can be written:  $V_i = E_x.h_{ix} + E_y.h_{iy}$, where $h_{ix}$ and $h_{iy}$ are the projections of the $\vec{h}_i$ antenna on the $\vec{x}$ and $\vec{y}$ axes. The combination of autocorrelations are then:
\begin{align}
A_++A_- =& \left<E_xE_x^*\right>\left(h_{1x}^2 + h_{2x}^2\right) + \left<E_yE_y^*\right>\left(h_{1y}^2 + h_{2y}^2\right)\nonumber\\
&+ \left(\left<E_xE_y^*\right>+\left<E_yE_x^*\right>\right)\left(h_{1x}h_{1y} + h_{2x}h_{2y}\right)\nonumber\\
A_+-A_- =& \left(h_{2x}h_{1y} - h_{1x}h_{2y}\right)\left(\left<E_xE_y^*\right>-\left<E_yE_x^*\right>\right)\nonumber
\end{align}
In the wave frame, the Stokes parameters are related to the electric field component auto and cross correlations as: $\left<E_xE_x^*\right>=SZ_0(1+Q)$, $\left<E_yE_y^*\right>=SZ_0(1-Q)$, $\left<E_xE_y^*\right>=SZ_0(U-iV)$ and $\left<E_yE_x^*\right>=SZ_0(U+iV)$, where $Z_0$ is the impedance of free space. The combination of autocorrelations are thus:
\begin{align}
A_++A_- =& SZ_0\left[ (1+Q)\left(h_{1x}^2 + h_{2x}^2\right) + (1-Q)\left(h_{1y}^2 + h_{2y}^2\right) + 2U\left(h_{1x}h_{1y} + h_{2x}h_{2y}\right)\right]\nonumber\\
A_+-A_- =& -2iV\,SZ_0\left(h_{2x}h_{1y} - h_{1x}h_{2y}\right)\nonumber
\end{align}
If the observed radio wave is purely circularly polarized (i.e., $U=0$ and $Q=0$), $(A_++A_-)/2$ provides $S$, after correction for the antenna orientation in the wave frame, and $(A_++A_-)/2i$ provides the circular polarization degree of the wave.

\begin{figure}
\includegraphics[width=\linewidth]{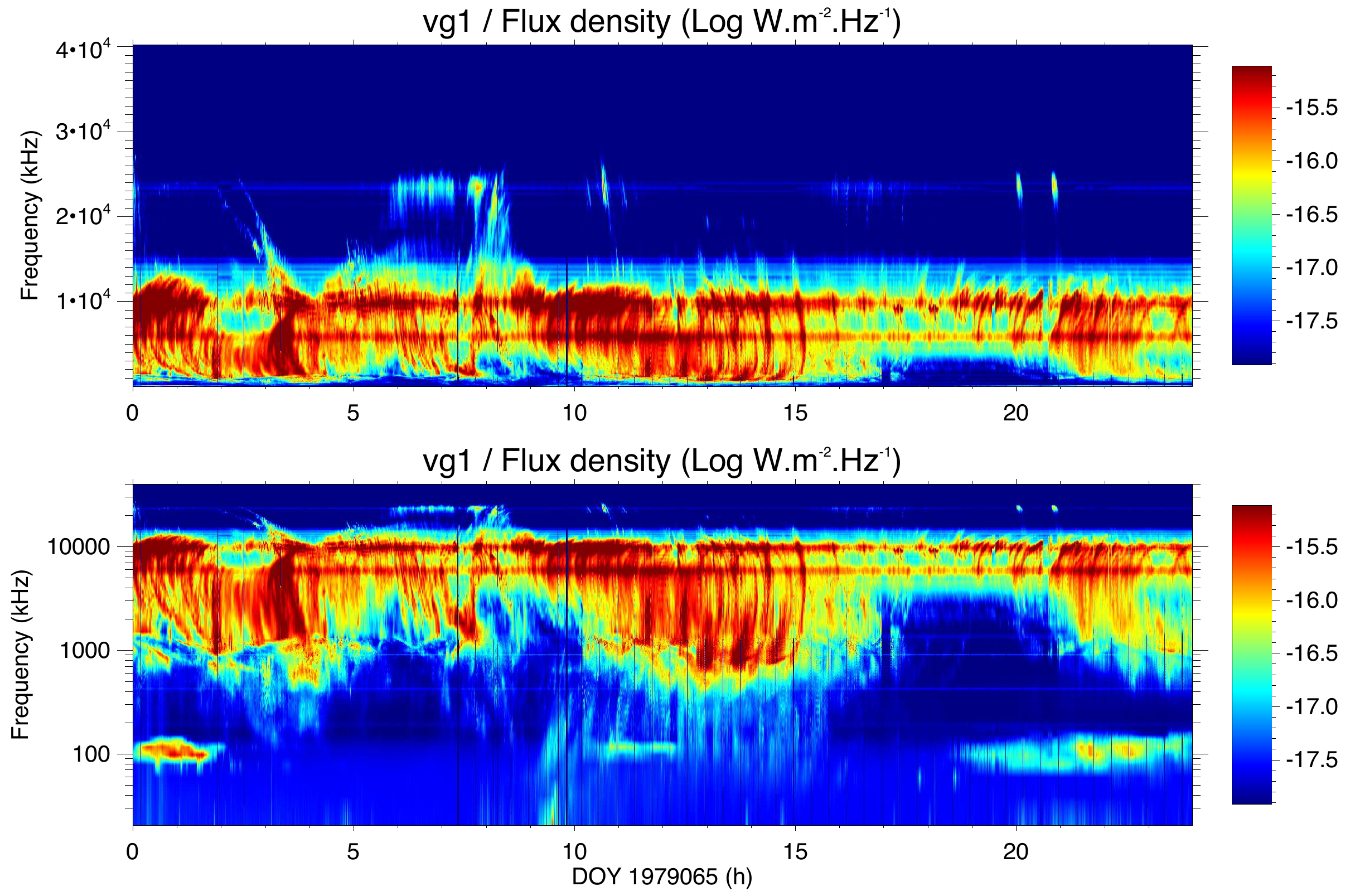}
\caption{Level 2 data summary plot for Voyager 1 during its Jupiter flyby, on 1979, March 6th. The upper plot shows the raw spectral flux density with a linear frequency axis, whereas the lower one shows the same data with a log-scale frequency axis.}\label{fig:l2}
\end{figure}

\begin{figure}
\includegraphics[width=\linewidth]{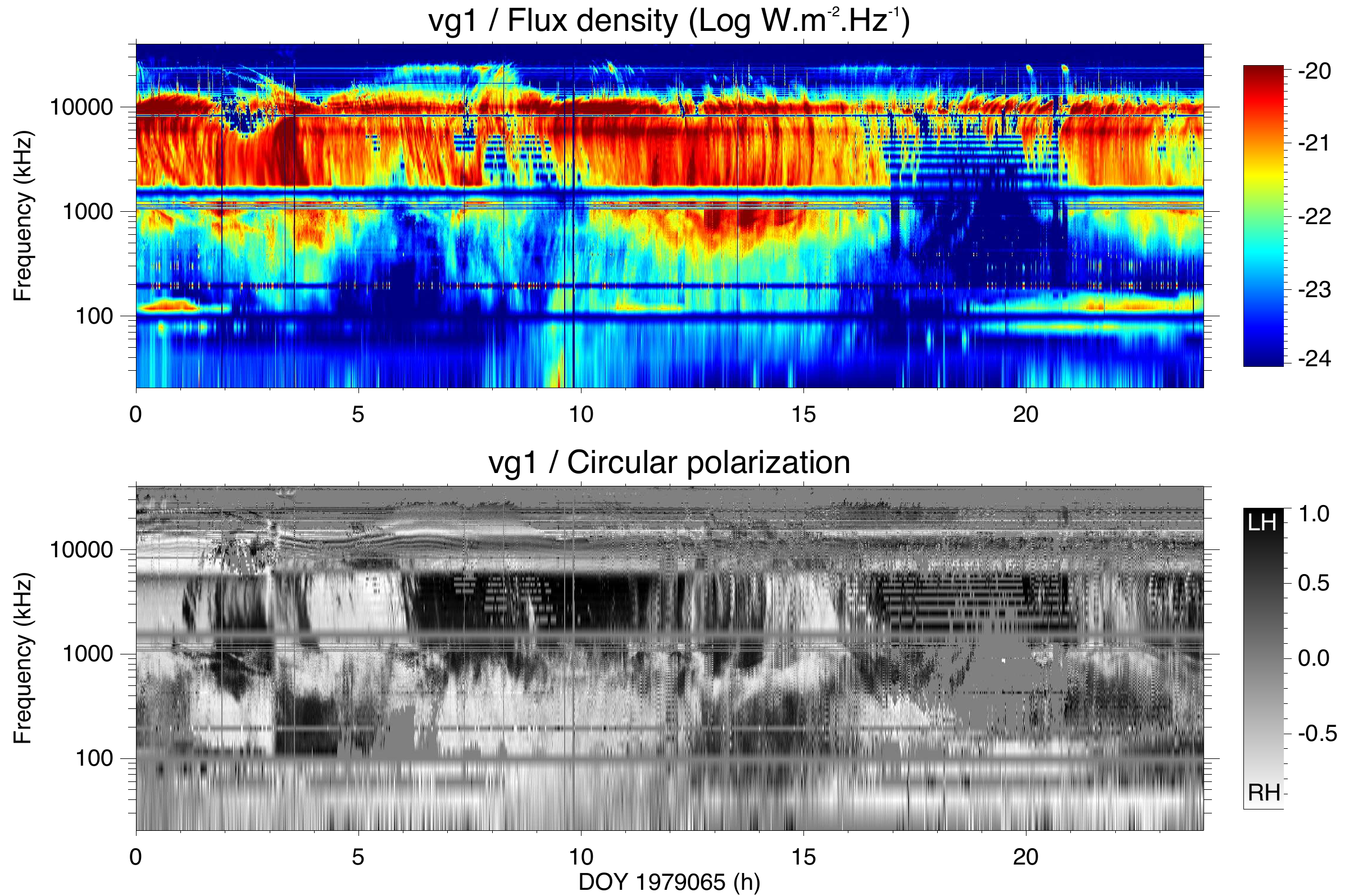}
\caption{Level 3f data summary plot for Voyager 1 during its Jupiter flyby, on 1979, March 6th, using the physical antenna parameters. The upper plot shows the spectral flux density with a log-scaled frequency axis, whereas the lower one shows circular polarization degree with the same frequency axis.}\label{fig:l3f}
\end{figure}

\begin{figure}
\includegraphics[width=\linewidth]{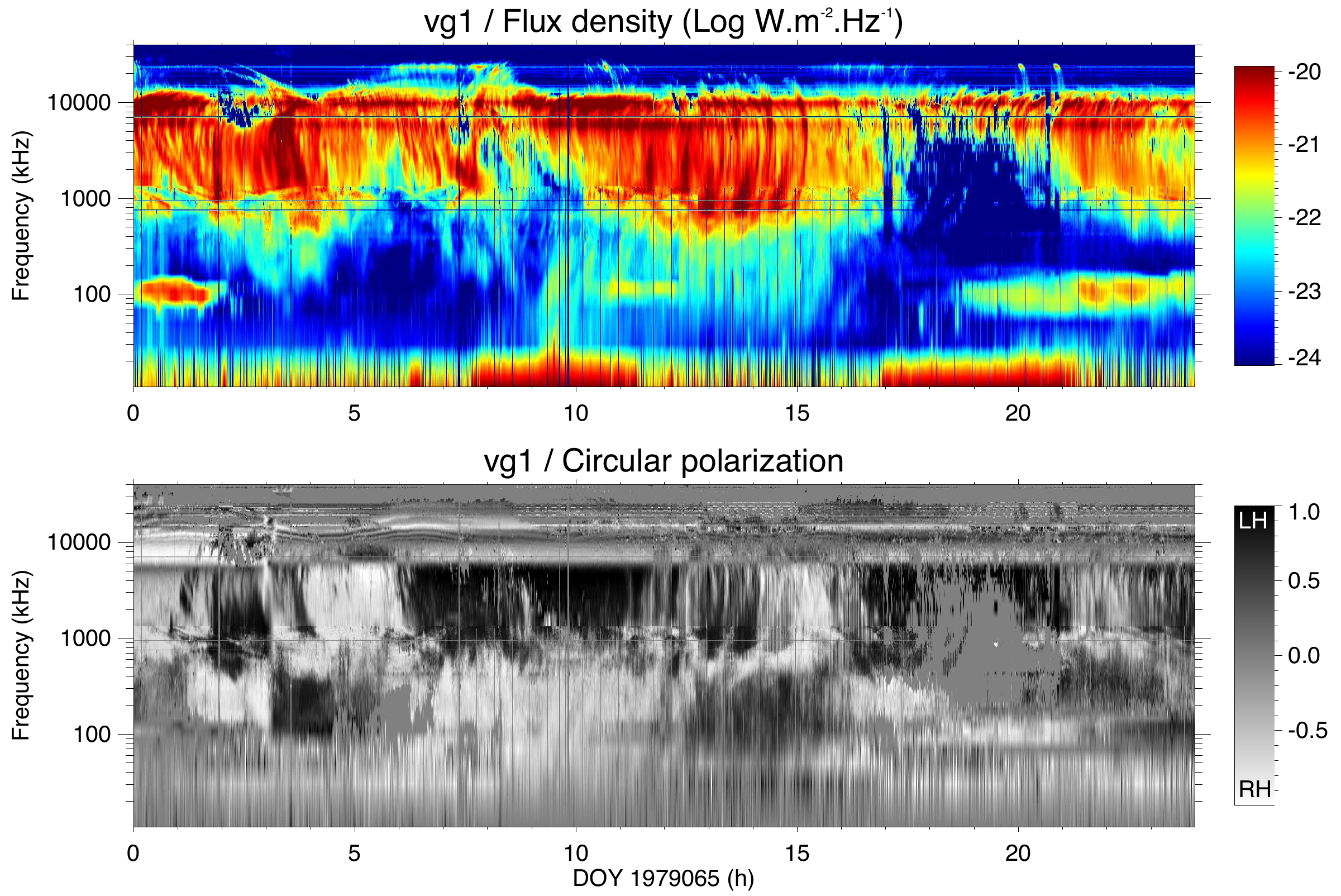}
\caption{Level 3t data summary plot for Voyager 1 during its Jupiter flyby, on 1979, March 6th, using the physical antenna parameters. The upper plot shows the spectral flux density with a log-scaled frequency axis, whereas the lower one shows circular polarization degree with the same frequency axis.}\label{fig:l3t}
\end{figure}

The derivation of the wave Stokes parameters requires a good calibration of the antenna effective lengths and directions. Several authors have published different calibrations for the PRA antenna. The physical antenna parameters (actual directions and lengths of the sensors) are provided in [Warwick et al. 1977]. The first sets of effective antenna parameters were published in [Ortega--Molina and Daigne, 1984] and [Ortega--Molina and Lecacheux, 1990]. A more recent study explores asymmetric antenna patterns [Wang and Carr, 1994]. Applying these effective antenna parameters on the L3f and L3t inversion will be conducted in the coming months. Figures \ref{fig:l2}, \ref{fig:l3f} and \ref{fig:l3t} show the measured flux density and the circular polarization degrees using the two inversions. The data presented in Figures \ref{fig:l3f}, and \ref{fig:l3t} use physical antenna parameters for the goniopolarimetric inversions. A background is removed from the L2 data before processing the L3f and L3t data. The vertical polarization reversal at about $03:00$ SCET shows that the antenna directions are not correctly handled. This is expected as the effective antenna directions may be very different from the physical antenna directions. 

\section{Datasets and Access}
The refurbished Voyager PRA full resolution data collection is composed of 5 datasets: 
Level 1 dataset (temporally sorted DEDR scans re-dispatched into daily files and operating modes); Level 2 dataset (daily files of ``POLLO'' mode DEDR individual records); 
Level 3f dataset (daily files of flux and polarization derived from ``POLLO'' mode DEDR Level 2 data, using the L3f inversion); Level 3t dataset (daily files of flux and polarization derived from ``POLLO'' mode DEDR Level 2 data, using the L3t inversion); Ephemeris dataset (daily files of planetary ephemeris, one file per planet, in the spacecraft frame); Summary plot dataset (daily summary plots for L2, L3f and L3t data, as shown on Figures \ref{fig:l2}, \ref{fig:l3f} and \ref{fig:l3t}).

Since this work is still in progress, data are not publicly available, yet. When completed, the data collection will be available from the MASER (Measurement, Analysis and Simulations of Emission in the Radio range) service\footnote{MASER web site: \url{http://maser.lesia.obspm.fr}} at the Observatoire de Paris. Direct access to data for download will be available at: \url{http://maser.obspm.fr/data/voyager/pra}. Final data products will be prepared in CDF (Common Data Format), so that common tools (like Autoplot [Faden et al, 2010]) can handle the data easily. The data products will also be available from the VESPA (Virtual European Solar and Planetary Access) [Erard et al, 2017] query portal\footnote{VESPA query portal: \url{http://vespa.obspm.fr}}.

\section{Discussion and further Work}
The two goniopolarimetric inversions that we have set up are complementary. The L3f inversion focusses on data measured at the same frequency, but on consecutive sweeps (i.e., the individual samples of each pair of L2 data are measured 6 seconds appart). The L3t inversion conversely focusses on data measured almost simultaneously in adjacent frequency bins. L3f data are more subject to interference on flux measurements, since data are extracted from a single spectral bin. As seen in Figure \ref{fig:l3f}, interference with constant power (removed with the background)  produce spectral gaps (as seen at about 100 kHz and 1.5 MHz), or background spectral lines (see between 2 and 10 MHz).   L3f data are more immune to interference lines because they use measurements from adjacent spectral bins. This produces, however, a broadening of spectral feature, as seen in Figure \ref{fig:l3t}, especially for nKOM emissions around 100 kHz and at the lowest frequencies below 30 kHz. The same remarks apply for polarization measurements. The full characterization of the L3t and L3f inversion sensitivity to interference and signal fluctuations will be studied in the next months. However, the next step will be to compare and select the calibrated electric antenna parameters that have been previously published.

The reprocessing of the Voyager PRA data will provide the community with a unique dataset. The preliminary data samples presented in this paper clearly show that the reanalysis of this dataset will be very interesting, especially in light of the discoveries by the Galileo, Cassini and Juno spacecraft at Jupiter and Saturn. At Jupiter, we will search radio emissions modulated by the Galilean satellites, including the study of the polarization of the corresponding radio bursts [Louis et al., this issue]. At Saturn, the high temporal resolution measurements with polarization will provide crucial clues and new ideas on the still unexplained rotational modulation of the kronian system [Lamy et al, 2011].

\section{Acknowledgments}
The authors thank F. Bagenal (LASP, Boulder, Colorado) for helping us hunt down documentation sources. They also thank A. Ortega-Molina for thoughtful and helpful comments during the preparation of the manuscript. 

\section*{References}
\everypar={\hangindent=1truecm \hangafter=1}

Burke,~B.\,F., and K.\,L.~Franklin, Radio emission from Jupiter, \textsl{Nature}, \textbf{175}, 1074, 1955.

Cecconi,~B., \'Etude goniopolarim\'{e}trique des \'{e}missions radio
de Jupiter et Saturne \`{a} l'aide du r\'{e}cepteur radio de la
sonde Cassini, PhD Thesis, University of Paris, Meudon, France,
April 2004.

Cecconi,~B., Influence of an extended source on goniopolarimetry (or direction finding) with Cassini and Solar TErrestrial RElations Observatory radio receivers, \textsl{Radio Sci.}, \textbf{42}, RS2003, 2007.

Cecconi,~B., and P.~Zarka, Direction finding and antenna calibration through analytical inversion of radio measurements performed using a system of two or three electric dipole antennas on a three-axis stabilized spacecraft, \textsl{Radio Sci.}, \textbf{40}, RS3003, 2005.

Erard,~S., B.~Cecconi, P.~Le\,Sidaner, A.\,P.~Rossi, M.\,T.~Capria, B.~Schmitt, V.~G\'enot, et al, VESPA: a Community-Driven Virtual Observatory in Planetary Science, \textsl{Planet. Space Sci.}, submitted, 2017.

Lamy,~L., Variability of southern and northern SKR periodicities, in \textsl{Planetary Radio Emissions VII}, edited by H.\,O.~Rucker, W.\,S.~Kurth, P.~Louarn, and G.~Fischer, Austrian Academy of Sciences Press, Vienna, 39--50, 2011.

Faden,~J., R.\,S.~Weigel, J.~Merka, and R.\,H.\,W.~Friedel, Autoplot: a Browser for Scientific Data on the Web. \textsl{Earth Sci. Inform.}, \textbf{3}, 41--49, doi:10.1007/s12145-010-0049-0, 2010.

Huc,~C., Voyager, Sauvetage des Donn\'ees de l'Exp\'erience PRA, \textsl{PLAS-HIS-VOYAGE\_PRA-00181-CN}, CNES, 1996.

Ortega--Molina,~A. and G.~Daigne, Polarization response of two crossed monopoles on a spacecraft, \textsl{Astron. Astrophys.}, \textbf{130}, 301--310, 1984.

Ortega--Molina,~A., and A.~Lecacheux, Polarization response of the Voyager-PRA experiment at low frequencies, \textsl{Astron. Astrophys.}, \textbf{229}, 558--568, 1990.

Street, D.~J., Voyager, Experiment Data Record Format Specification. \textsl{618-306 Rev.\ D}, NASA-JPL, 1977.

Wang,~L., and T.\,D.~Carr, Recalibration of the Voyager PRA antenna for polarization sense measurement, \textsl{Astron. Astrophys.}, \textbf{281}, 945--954, 1994.

Warwick,~J.\,W., J.\,B.~Pearce, R.\,G.~Peltzer, and A.\,C.~Riddle, Planetary Radio Astronomy experiment for Voyager missions, \textsl{Space Sci. Rev.}, \textbf{21}, 309--327, 1977.

Warwick, J. W., J.\,B.~Pearce, A.\,C.~Riddle, J.\,K.~Alexander, M.\,D.~Desch, M.\,L.~Kaiser, J.\,R.~Thieman, T.\,D.~Carr, S.~Gulkis, A.~Boischot, C.\,C.~Harvey, and B.\,M.~Pedersen, Voyager--1 Planetary
Radio Astronomy observations near Jupiter, \textsl{Science}, \textbf{204}, 995--998, 1979a.

Warwick,~J.\,W., J.\,B.~Pearce, A.\,C.~Riddle, J.\,K.~Alexander, M.\,D.~Desch, M.\,L.~Kaiser, J.\,R.~Thieman, T.\,D.~Carr, S.~Gulkis, A.~Boischot, Y.~Leblanc, B.\,M.~Pedersen, and D.\,H.~Staelin, Planetary Radio Astronomy observations from Voyager~2 near Jupiter, \textsl{Science}, \textbf{206}, 991--995, 1979b.

Warwick,~J.\,W., J.\,B.~Pearce, D.\,R.~Evans, T.\,D.~Carr, J.\,J.~Schauble, J.\,K.~Alexander, M.\,L.~Kaiser, M.\,D.~Desch, B.\,M.~Pedersen, A.~Lecacheux, G.~Daigne, A.~Boischot, and C.\,H.~Barrow,
Planetary Radio Astronomy observations from Voyager~1 near Saturn, \textsl{Science}, \textbf{212}, 239--243, 1981.

Warwick,~J.\,W., D.\,R.~Evans, J.\,H.~Romig, J.\,K.~Alexander, M.\,D.~Desch, M.\,L.~Kaiser, M.~Aubier, Y.~Leblanc, A.~Lecacheux, and B.\,M.~Pedersen, Planetary Radio astronomy Observations from Voyager~2 near Saturn, \textsl{Science}, \textbf{215}, 582--587, 1982.

Warwick, J. W., et al. (16 co--authors), Voyager~2 radio observations of Uranus, \textsl{Science}, \textbf{233}, 102--106, 1986.

Warwick, J. W., et al. (22 co--authors), Voyager planetary radio astronomy at Neptune, \textsl{Science}, \textbf{246}, 1498, 1989.

Zarka,~P., Auroral radio emissions at the outer planets: Observations and theories, \textsl{J. Geophys. Res.}, \textbf{103}, 20159--20194, 1998.

Zarka,~P., and B.\,M.~Pedersen, Statistical study of Saturn electrostatic discharges, \textsl{J. Geophys. Res.}, \textbf{88}, 9007--9018, 1983.

Zarka,~P., and B.\,M.~Pedersen, Radio detection of Uranian lightning by Voyager~2, \textsl{Nature}, \textbf{323}, 605--608, 1986.
\end{document}